# Transportation Cyber Incident Awareness through Generative AI-Based Incident Analysis and Retrieval-Augmented Question-Answering Systems

**Ostonya Thomas\***
M.S. Student, Glenn Department of Civil Engineering
Clemson University, Clemson, South Carolina, 29634
Email: ostonyt@clemson.edu

**Muhaimin Bin Munir**
Research Assistant
Data Mining Lab
The University of Texas at Dallas,
800 W Campbell Rd, Richardson, TX 75080
Email: muhaimin.binmunir@utdallas.edu

**Jean-Michel Tine**
Ph.D. Student, Glenn Department of Civil Engineering
Clemson University, Clemson, South Carolina, 29634
Email: jtine@clemson.edu

**Mizanur Rahman, Ph.D.**
Assistant Professor, Department of Civil, Construction & Environmental Engineering
The University of Alabama, Tuscaloosa, Alabama, 35487
Email: mizan.rahman@ua.edu

**Yuchen Cai**
Research Assistant
Data Mining Lab
The University of Texas at Dallas,
800 W Campbell Rd, Richardson, TX 75080
Email: yuchen.cai@utdallas.edu

**Khandakar Ashrafi Akbar**
Research Assistant
Data Mining Lab
The University of Texas at Dallas,
800 W Campbell Rd, Richardson, TX 75080
Email: ashrafi@utdallas.edu

**Md Nahiyan Uddin**
Research Assistant
Data Mining Lab
The University of Texas at Dallas,
800 W Campbell Rd, Richardson, TX 75080
Email: md.uddin@utdallas.edu

**Latifur Khan, Ph.D.**
Professor
Department of Computer Science
The University of Texas at Dallas,
800 W Campbell Rd, Richardson, TX 75080
Email: lkhan@utdallas.edu

**Trayce Hockstad, J.D., M.A.**
Law & Policy Analyst, Transportation Policy Research Center
The University of Alabama, Tuscaloosa, Alabama, 35487
Email: tahockstad@ua.edu

**Mashrur Chowdhury, Ph.D., P.E.**
Eugene Douglas Mays Chair of Transportation, Glenn Department of Civil Engineering
Clemson University, Clemson, South Carolina, 29634
Email: mac@clemson.edu

**ABSTRACT**

Technological advancements have revolutionized numerous industries, including transportation. While digitalization, automation, and connectivity have enhanced safety and efficiency, they have also introduced new vulnerabilities. With 95% of data breaches attributed to human error, promoting cybersecurity awareness in transportation is increasingly critical. Despite numerous cyberattacks on transportation systems worldwide, comprehensive and centralized records of these incidents remain scarce. To address this gap and enhance cyber awareness, this paper presents a large language model (LLM)-based approach to extract and organize transportation-related cyber incidents from publicly available datasets. A key contribution of this work is the use of generative AI to transform unstructured, heterogeneous cyber incident data into structured formats. Incidents were sourced from the Center for Strategic & International Studies (CSIS) List of Significant Cyber Incidents, the University of Maryland Cyber Events Database (UMCED), the European Repository of Cyber Incidents (EuRepoC), the Maritime Cyber Attack Database (MCAD), and the U.S. DOT's Transportation Cybersecurity and Resiliency (TraCR) Examples of Cyber Attacks in Transportation (2018 to 2022). These were classified by a fine-tuned LLM into five transportation modes: aviation, maritime, rail, road, and multimodal, forming a transportation-specific cyber incident database. Another key contribution of this work is the development of a Retrieval Augmented Generation question-answering system, designed to enhance accessibility and practical use by enabling users to query the curated database for specific details on transportation-related cyber incidents. By leveraging LLMs for both data extraction and user interaction, this study contributes a novel, accessible tool for improving cybersecurity awareness in the transportation sector.

**INTRODUCTION**

The transportation industry has become more digitalized, automated and connected than ever before. While this is beneficial to safety and efficiency, this expansion of cyberspace has dramatically increased the attack surfaces in the field. The 2024 IBM Cost of a Data Breach Report reported that the average cost of a data breach in transportation in 2024 was $ 4.43 million (*1*). In 2020, it was reported that the maritime industry experienced a 900% influx in cyberattacks over the previous three years (*2*). The Chief Information Security Officer of the Port of Los Angeles reported that the port experienced a significant increase in cyberattacks from 7 million per month in 2014 to a whopping 60 million attacks per month in 2024 (*3*). In 2024, there was reported to have been a 220% increase in rail-related cyberattacks over the previous five years (*4*). The automotive industry has been reported to have experienced a 225% increase in cyberattacks over the past three years as of 2025 (*5*). A report from 2025 indicated that the aviation industry experienced a 600% year-on-year increase in ransomware attacks (*6*).

Cyberattacks have been increasing in volume and sophistication (*7*). However, it must not be overlooked that 95% of data breaches are a result of human error (*8*). As systems become more advanced, humans must work closely with intelligent systems; consequently, insufficiently trained staff pose a serious cybersecurity risk, as they are often exploited by attackers through social engineering (*9*). It follows that one way to improve cybersecurity in the transportation sector is through employee education, thus enhancing the cyber awareness of the workforce (*10*). One means of employee education is through ensuring employees have access to up-to-date information regarding cyber incidents that have occurred in the transportation sector over the years. However, it is challenging to find databases or lists of cyber incidents in transportation spanning over multiple years.

Over the last decade, a variety of publicly available repositories have aimed to catalog cyber incidents across industries, including transportation. For instance, the Cyber Events Database at the University of Maryland assembles reported attacks dating back to 2014, providing structured incident summaries drawn from news reports and press releases (*11*). The European Repository of Cyber Incidents (EuRepoC) similarly offers an open-access, daily updated log of global events, although its focus spans all sectors rather than transportation specifically (*12*). While these resources deliver valuable archives, they typically require manual browsing or custom data-export scripts to locate incidents relevant to a particular domain or question. In parallel, vulnerability trackers such as the National Vulnerability Database (NVD) index software flaws and exploits but do not directly connect them to real-world transportation cyber-incidents (*13*). Transportation-focused guidance from agencies such as the Cybersecurity and Infrastructure Security Agency (CISA) and the European Union Agency for Cybersecurity (ENISA) offers best practices and sector-wide threat assessments (*14, 15*). However, it lacks an incident-level search capability that can answer natural-language queries like "What has happened to transit agencies?" or "Which freight operators suffered ransomware attacks?".

LLMs have previously been used for text classification and database development tasks. For example, Kumar et al. (*16*) fine-tuned GPT and LLaMA-2 models and compared their performance across a range of text-classification tasks, concluding that both models demonstrated strong capabilities in this domain. Similarly, Berkane et al. (*17*) developed an end-to-end, human-in-the-loop LLM application that utilizes textual descriptions to construct a target dataset via web crawling, illustrating how LLMs can aid in database construction. Our approach also employs an LLM, specifically a fine-tuned GPT-3.5 Turbo model, to classify the transportation mode affected in cyber incidents for database generation. However, unlike

Berkane et al. (*17*), we do not rely on web crawling; instead, we filter incidents from reliable cybersecurity-specific data sources.

LLMs have also been applied across various cybersecurity tasks. Xu et al. (*18*) conducted a systematic literature review of 185 papers on LLM use in cybersecurity, outlining applications in network, information, software, hardware, and blockchain security. Their study highlights the frequent use of open-source, collected, and constructed datasets, as well as common techniques such as retrieval-augmented generation (RAG), prompt engineering, and fine-tuning. The review also noted the predominant use of decoder-only models such as GPT-2, GPT-3.5, and GPT-4. However, Xu et al. (*18*) did not identify LLM applications specifically designed for the transportation sector or cybersecurity education. In contrast, Salek et al. (*19*) integrated LLMs with RAG and fine-tuning within a threat modeling framework to map attack paths in transportation cyber-physical systems while minimizing expert intervention. Our approach differs in that it is designed to support user learning and awareness rather than reduce human involvement in threat analysis.

Educational applications of LLMs in cybersecurity are also emerging. For instance, CyberRAG (*20*) offers an interactive Q&A system that supports cybersecurity education using ontology-aware RAG. Similarly, IntellBot (*21*) utilizes LangChain and RAG, drawing from a knowledge base comprising reports, security books, blogs, and the National Vulnerability Database, to create a chatbot that helps users proactively guard against cyber threats. While these systems share educational goals with our work, our approach is uniquely focused on the transportation domain and specifically designed to answer user questions about cyber incidents affecting transportation systems.

This study is inspired by such prior work in LLM-based database construction and natural language querying. Building on approaches like those of Kumar et al. and Berkane et al., it introduces three key innovations. First, it targets the transportation sector, which is underrepresented in LLM-enabled cybersecurity research. Second, it forgoes web crawling in favor of verified, domain-specific, heterogenous data sources. To harmonize these diverse formats, we employed generative AI with carefully crafted one-shot prompt designs to convert the data into a uniform JSON structure. Third, it is not just educational in nature but purpose-built for helping users explore historical cyber incidents in transportation.

Our work makes two core contributions. The first is the creation of a structured, mode-specific database of transportation-related cyber incidents, built by extracting and classifying entries from trusted cybersecurity datasets using a fine-tuned GPT-3.5 Turbo model. The second is the development of a question-answering (Q-A) system using RAG and GPT-4o mini, which leverages this static database to respond to user queries. **Figure 1** illustrates how these two components are connected.

By converting heterogeneous data into a uniform format and compiling each entry with key details such as attack type, date, transportation mode, and a brief description, our system serves as a transportation-specific cyber incident database and provides users with precise, contextual summaries. For example, a query like "How much ransom was demanded from Kia Motors America in the 2022 cyberattack?" yields a targeted response drawn directly from the curated data, answering the question while also providing relevant contextual details. Emphasizing accuracy, domain specificity, and user awareness, our system fills a gap in existing tools without offering real-time alerts or forensic analysis.

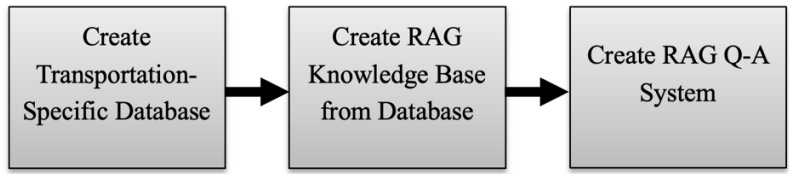

**Figure 1 Process flow from database creation to user Q-A system**

**LLM-BASED CYBER INCIDENT INFORMATION EXTRACTION APPROACH**

This LLM-based cyber incident information extraction approach enables individuals in the transportation industry to learn more about cyberattacks that have occurred in their respective modes of transportation. Utilizing existing databases, it filters out incidents unrelated to transportation to create a database of transportation-related cyber incidents. The purpose of this tool is to enable users to become more aware of the cyber threat landscape in the transportation industry as a whole.

**Existing Datasets**

Five data sources were used for this study, the University of Maryland Cyber Events Database (UMCED) (*11*), European Repository of Cyber Incidents (EuRepoC) Database (*12*), Center for Strategic and International Studies (CSIS) List of Significant Cyber Incidents (*22*), NHL Stenden Maritime Cyber Attack Database (MCAD) (*23*), and the Transportation Cybersecurity and Resiliency (TraCR) Examples of Cyber Attacks in Transportation from 2018 to 2022,  a manually crafted dataset of cyber incidents in transportation from 2018 to 2022 (our database). The latter two data sources only contain transportation-related cyber incidents, while the former sources contain cyber incidents related to key sectors (including transportation). The UMCED contains publicly available information related to cyber incidents from 2014 to the current day. This database was developed by the University of Maryland's Center for International and Security Studies at Maryland. It was created to provide structured data to inform decisions regarding investments in cyber incident prevention and response (*11*). The EuRepoC database contains incidents from 2000 to the current day. This database was compiled for the scientific analysis of cyber events, enabling a better understanding of the current cyber threat landscape (*12*). The CSIS data source is a list of significant cyber incidents that have occurred since 2006. This list contains descriptions of attacks on "government agencies, defense and high-tech companies, or economic crimes with losses of more than a million dollars" (*22*). The MCAD has considered cyber incidents in the maritime sector since 2001. This database contains publicly available information for over 160 cyber incidents and aims to promote cybersecurity awareness in the maritime industry (*23*). The TraCR database contains cyber incidents that occurred in aviation, maritime, rail, and road transportation from 2018 to 2022. This data was collected from scholarly articles and cybersecurity reports to gain a deeper understanding of the threat landscape in transportation during the study period. **Table 1** summarizes the datasets used for this study and the information they offer.

**TABLE 1 Overview of Cyber Incident Databases Utilized to Construct the Transportation-Specific Database for this Study**

| Database | Description | Number of Incidents at Time Collection | Types of Incidents | Reference |
|---|---|---|---|---|
| **UMCED** | Publicly available information related to cyber incidents from 2014 to the current day | 14,041 | Multiple industries and sectors | (*11*) |
| **EuRepoC** | Cyber incidents from 2000 to the current day compiled for the scientific analysis of cyber events | 3,414 | Multiple industries and sectors | (*12*) |
| **CSIS** | Descriptions of attacks on "government agencies, defense and high-tech companies, or economic crimes with losses of more than a million dollars" since 2006 | (Not specified, 97 pages of incidents) | Multiple industries and sectors | (*22*) |
| **TraCR** | Examples of cyber incidents related to transportation that occurred in aviation, maritime, rail, and road transportation from 2018 to 2022 | 124 | Transportation industry only (aviation, maritime, rail, road and multimodal) | |
| **MCAD** | Publicly available information regarding cyber incidents in the maritime sector since 2001 | 279 | Maritime only | (*23*) |

**Database Preparation**

This study used data from the five data sources discussed above. As aforementioned, of the five datasets, two contained cyber incidents solely related to transportation, and the three others contained a mixture of incidents related to key sectors. Since these data sources are formatted differently, each dataset was converted to a JSON file. **Figure 2** demonstrates how the data sources were organized into JSON format. The TraCR and CSIS datasets were PDF files, so GPT-4o was used to extract relevant fields from the incidents and assign a transportation mode to each incident to form structured entries into JSON files, creating one file per dataset (seen in **Figure 2 (a)**). The UMCED dataset was available in an Excel file, the EuRepoC file was available in a CSV file, and the MCAD was scraped into a structured format as a text file. As a result, those datasets were easily converted to JSON format without the use of the LLM (seen in **Figure 2 (b)**).

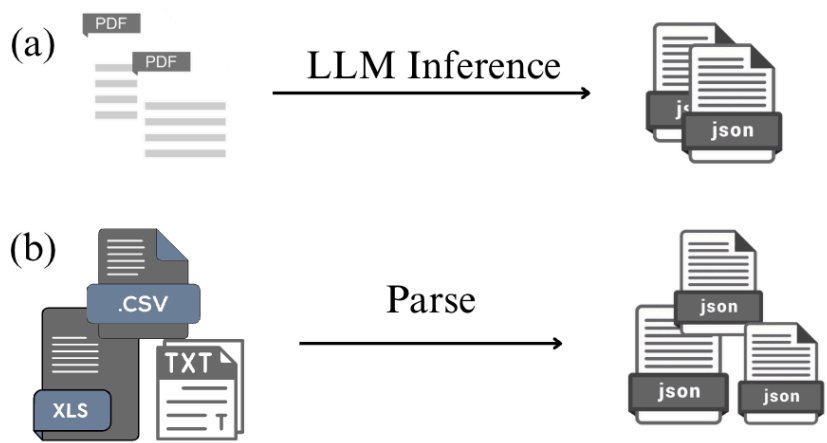

**Figure 2 (a) PDF datasets converted using GPT-4o to JSON; (b) CSV, XLS, and TXT files parsed to JSON without LLM involvement**

Since the TraCR dataset already had 124 examples of cyber incidents in transportation (115 entries, since some entries referred to multiple incidents) with their specific modes, these entries along with 85 examples of incidents not related to transportation from the UMCED, CSIS, and EuRepoC data sources served as a training set to fine-tune a GPT 3.5 Turbo model to classify the transportation mode of cyber incidents correctly. To assign a transportation mode label to each incident, we used a single-turn prompt that instructed the model to classify the mode of transportation involved based on incident data found in the JSON object. The prompt included key fields from the structured entry, such as attack name, incident type, description, victim, attacker, and motive. The model was then asked to populate the transportation mode field using predefined guidelines. These guidelines mapped domain-specific terms to one of five categories: Road, Rail, Maritime, Aviation, or Multimodal. If no transportation-related elements were present, the model was instructed to leave the field unpopulated (null). The fine-tuned model was trained on 200 incidents for two epochs.

Using the fine-tuned LLM, the UMCED, CSIS, and EuRepoC datasets were assigned transportation modes based on their descriptions. To evaluate the model, five incidents per transportation mode ("Aviation," "Maritime," "Multimodal," "Rail," "Road," and "None") per dataset were selected using a random seed. The evaluation set had a total of 90 randomly selected incidents. The retained incidents from the filter along with the TraCR and MCAD incidents were converted to vector store embeddings to serve as a knowledge base for the Q-A system. **Figure 3** summarizes the database preparation process and **Figure 4** shows an example of a retained incident from the CSIS dataset.

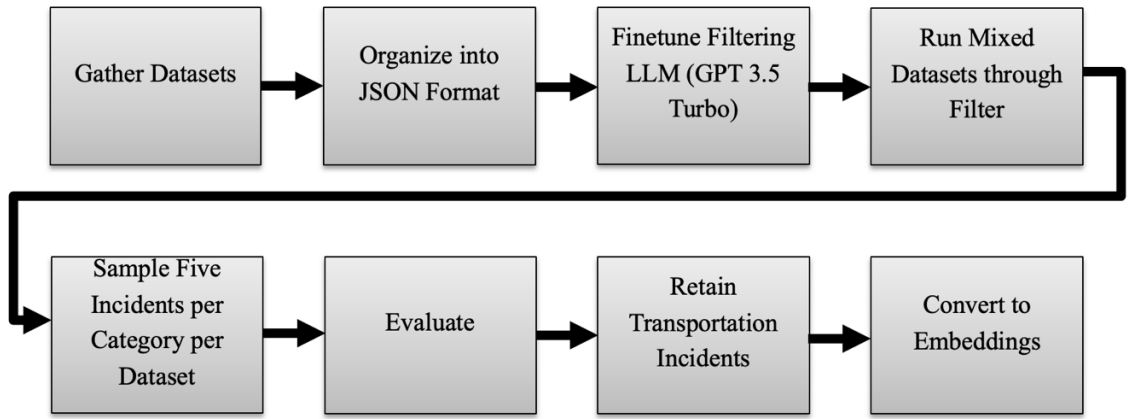

**Figure 3 Database preparation process flow**

```
{
  "attack_name": "Chinese hackers exfiltrate United Airlines data",
  "incident_type": "Data Breach",
  "description": "Chinese hackers exfiltrated significant amounts of customer data from United Airlines.",
  "Date": "May 2015",
  "detection": null,
  "victim": {
  "name": "United Airlines",
  "country": "USA",
  "category": "corporate"
  },
  "attacker": {
  "name": "Chinese hackers",
  "country": "China",
  "category": "state institution"
  },
  "Motive": "financial",
  "database_entry_date": null,
  "Reference": null,
  "Transportation_mode": "Aviation"
  },
```

**Figure 4 Example of a retained incident entry from the CSIS dataset after being assigned a transportation mode by the filter**

## RETRIEVAL AUGMENTED GENERATION-BASED QUESTION-ANSWERING APPROACH

The database described above serves as the knowledge base for the Q-A approach. Its textual content was processed using LlamaIndex, which converted the documents into embeddings via OpenAI's text-embedding-ada-002 model. These embeddings were stored in a persistent vector store, allowing the construction of a vector index for efficient similarity-based retrieval. Chunks of 768 tokens with a 100-token overlap were used for creating the vector store index.

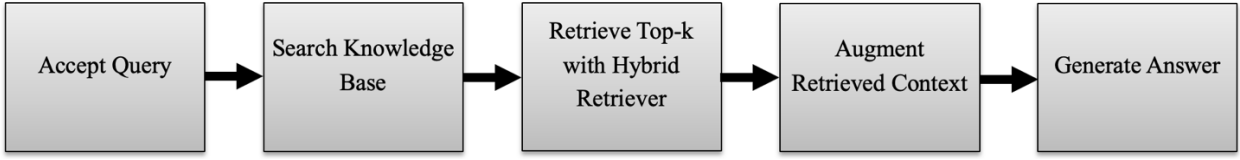

**Figure 5 RAG-based Q-A system using hybrid retrieval**

For the question answering task, RAG with ChatGPT-4o mini as the underlying model was used. RAG helps mitigate the issue of hallucinations (i.e., incorrect or fabricated responses sometimes produced by large language models) by retrieving the top $k$ most relevant chunks from the external knowledge base and incorporating them into the prompt. This is especially important in domains like cybersecurity, where LLMs may have limited knowledge and new incidents occur frequently. By grounding the model's responses in relevant domain-specific information, RAG helps ensure that users receive accurate and timely insights about the evolving threat landscape. **Figure 5** shows how our RAG Q-A system works.

Hybrid retrievers perform better at retrieval tasks than dense or sparse methods alone through capturing semantic similarity and lexical overlap (*24*, *25*). Our RAG system uses a hybrid retriever that uses dense and sparse retrieval. The dense retriever is based on vector similarity, while the sparse retriever uses BM25 (*26*). The dense retriever utilizes vector similarity between queries and chunks of information to identify the most relevant documents for each query. The similarity between a query $q$ and a chunk $c$ is calculated through cosine similarity:

$$\text{Vec}\,(q,c) = \cos\theta = \frac{\boldsymbol{q} \cdot \boldsymbol{c}}{\|\boldsymbol{q}\| \, \|\boldsymbol{c}\|} \tag{1}$$

BM25 (the sparse retriever) utilizes term frequency and inverse document frequency, with normalization applied to chunk length. The BM25 score for a query $q$ and a chunk $c$ is calculated as follows:

$$\text{BM25}\,(q,c) = \sum_{t \in q} \text{IDF}\,(t) \cdot \frac{f(t,c) \cdot (k_1 + 1)}{f(t,c) + k_1 \cdot \left(1 - b + b \cdot \frac{|c|}{\text{avgcl}}\right)} \tag{2}$$

where,
$f(t,c)$ is the frequency of term $t$ in chunk $c$,
$|c|$ is the length of a chunk $c$,
avgcl is the average chunk length in the knowledge base,
$k_1$ is the hyperparameter for term frequency (the default of 1.5 was used),
$b$ is the hyperparameter for chunk length normalization (the default of 0.75 was used), and
$\text{IDF}\,(t) = \log\left(\frac{N - n_t + 0.5}{n_t + 0.5} + 1\right)$, where $N$ is the total number of chunks and $n_t$ is the number of chunks containing term $t$.

A hybrid score for each chunk is calculated by:

$$\text{Score}(q, c_i) = \alpha \cdot \text{Vec}(q, c_i) + (1 - \alpha) \cdot \text{BM25}(q, c_i) \qquad (3)$$

where $\alpha \in [0,1]$ is a parameter that balances the influence of sparse and dense retrieval methods. For this experiment $\alpha$ was chosen to be 0.5 to give equal weight to semantic similarity and the sparse retrieval method. The top $k$ chunks for each retriever are retrieved and for the overlapping chunks, the hybrid scores are calculated. The chunks are ranked in descending order based on their hybrid scores and the top $k$ from that list serve as the context for the query.

To generate structured natural language answers, we designed a single-turn prompt template with clear formatting and relevant context tailored to the question-answering task instructing the LLM to assume the role of a cybersecurity analyst answering questions about transportation-related cyber incidents. To maximize the use of relevant information while staying within model input limits, the approach incorporates token-aware batching. This approach tracks the total token count while aggregating retrieved context passages and splits them into multiple batches when necessary. This enables the model to reason over more content without truncation. Additionally, the generation temperature was set to zero to ensure consistent and reproducible outputs. The prompt explicitly directed the model to respond using only the provided context without referring to it explicitly. To encourage faithful and detail-oriented responses, the instructions emphasized the inclusion of dates, affected locations, and impact summaries. When multiple context entries referenced the same incident (e.g., overlapping entities and timeframes), the model was asked to consolidate them into a unified answer while drawing directly from the descriptions with minimal rephrasing. Responses were constrained to professional, paragraph-style summaries with no citations, speculation, or source mentions. An example answer was provided in the prompt to anchor the expected format. Together, these components create a system that effectively balances accuracy, consistency, and context-awareness to deliver reliable answers about the evolving cyber threat landscape in transportation.

## EVALUATION

To assess the reliability and effectiveness of the system, the evaluation was conducted in two stages: database preparation and question-answering. The goal of the first stage was to determine whether the LLM could accurately extract and organize structured cyber incident data across different transportation modes. To test this, five incidents per transportation mode category were randomly sampled from each of the three filtered datasets using a fixed random seed. The second stage aimed to evaluate the ability of the LLM to provide accurate, relevant, and complete answers to user queries using the curated database. The question-answering system was evaluated by comparing the responses generated by the language model to the ground-truth answers for a set of 50 test questions. ROUGE-1, ROUGE-2, ROUGE-L, precision, accuracy, and recall were used as evaluation metrics. In addition, the Q-A responses were manually inspected to assess the factual accuracy and completeness of the LLM's answers beyond what automatic metrics could capture.

### Filter Evaluation

Thirty incidents were sampled from each of the three datasets, five from each transportation mode category. The performance of the database preparation LLM is summarized in **Table 2**. Upon manual inspection of the 90 samples from the CSIS, UMCED and EuRepoC datasets, it was found that the fine-tuned LLM was able to classify the transportation mode of incidents with an accuracy of 88.89%. While eight samples were incorrectly classified, it should be noted that six of those incidents were multimodal incidents in which the LLM only indicated one of the modes affected in the incident instead of classifying the incidents as multimodal. These incidents were only partially correctly classified. 82 incidents were correctly classified. One of the two completely incorrect incidents was an incident that affected the Naval Air Weapons Program. This was an aviation incident, but it was classified as a maritime incident. The other completely incorrect incident involved an attack on the International Marine Trainers' Association. This incident was unrelated to transportation but was classified as a maritime incident. These incidents may have been classified as related to maritime due to the presence of words consistently associated with the maritime industry such as "Naval" and "Marine." Notably, of the 15 incidents classified as unrelated to transportation, zero incidents were misclassified. From our sample, there were no instances of transportation-related incidents being classified as unrelated to transportation. Of the 75 sampled incidents classified as transportation-related and retained, 98.67% of the incidents (i.e., 74 incidents) were related to transportation. While this approach has successfully retained transportation incidents to form a database and knowledge base, it is worth noting that multiple datasets report the same incidents, resulting in a lack of a comprehensive number for individual incidents that have occurred in transportation. We do not claim to have captured all transportation-related cyber incidents that have ever occurred.

**TABLE 2 Filtering Approach Performance**

| Metric | Performance | Comments |
|---|---|---|
| **Accuracy** | 88.89% | Correct classification of 82/90 sampled incidents |
| **Incorrectly Labeled** | 8 incidents | Six misclassified incidents were multimodal incidents in which one of the modes was correctly classified, the other two incidents were misclassified entirely |
| **Correctly Labeled** | 82 incidents | High classification rate across modes |
| **False Nulls Sampled** | 0 incidents | Of the 15 sampled incidents classified as unrelated to transportation, zero incidents were misclassified |
| **Partially Correctly Labeled** | 6 incidents | One of the multiple modes was correctly classified; multi-modal classification needs to be improved |
| **Completely Incorrect** | 2 incidents | Both incidents were misclassified as impacting maritime, possibly due to the presence of words commonly associated with the maritime field |
| **Percentage of Retained Incidents Related to Transportation** | 98.67% | Of the 75 sampled incidents classified as impacting transportation, 74 were relevant to the transportation industry |

**Question-Answering Evaluation**

To evaluate the question-answering task, 10 incidents were selected from each of the five datasets, and a question was posed based on the information contained in the incident descriptions. For the sake of evaluation, these test questions were restricted to closed-ended questions. The ground-truths for the questions were constructed by combining the incident descriptions for each reported incident across all datasets. If an incident appeared in multiple datasets, their respective descriptions were merged into a single reference answer. These duplicates were identified manually by matching similar phrases, dates, and descriptions. However, this manual process may have led to some relevant entries being overlooked in the final ground-truths. **Figure 6** shows an example of an LLM response to a transportation-related cyber incident question with its ground-truth. For this study, ROUGE-1, ROUGE-2, and ROUGE-L scores and precision, accuracy and recall were used to evaluate the Q-A approach.

```
2. Q: How many of Air India's customers were impacted by the cyberattack in 2021?
A: In 2021, Air India experienced a data breach that impacted approximately 4.5 million of its
customers. This incident occurred two months after its passenger service system provider was
hacked, leading to the exposure of sensitive customer information. The breach involved the
personal data of customers who had traveled with Air India, although specific details regarding
the nature of the compromised data were not disclosed.

Ground Truth:
In 2021, 4.5 million of Air India's customers were impacted by a data breach two months after
its passenger service system provider was hacked.
```

**Figure 6 Example of Q-A LLM response to a question regarding a transportation-related incident followed by the ground-truth for the question**

ROUGE (*27*) is an evaluation metric that considers lexical overlap of n-grams, word sequences, and word pairs between generated text and reference text. Some variants include ROUGE-1 (unigram overlap), ROUGE-2 (bigram overlap), and ROUGE-L (longest common subsequence). For this study we use ROUGE-1, 2 and L as some of the evaluation metrics. **Figures 7, 8, and 9** show the ROUGE scores of the responses generated. The average ROUGE-1, ROUGE-2 and ROUGE-L scores were 0.54, 0.33, and 0.50 respectively. These results indicate that the model-generated answers captured a substantial portion of the relevant content (indicated by ROUGE-1 scores) and generally followed the structure of the reference summaries (indicated by ROUGE-L scores). The lower ROUGE-2 score suggests reduced alignment at the phrase level.

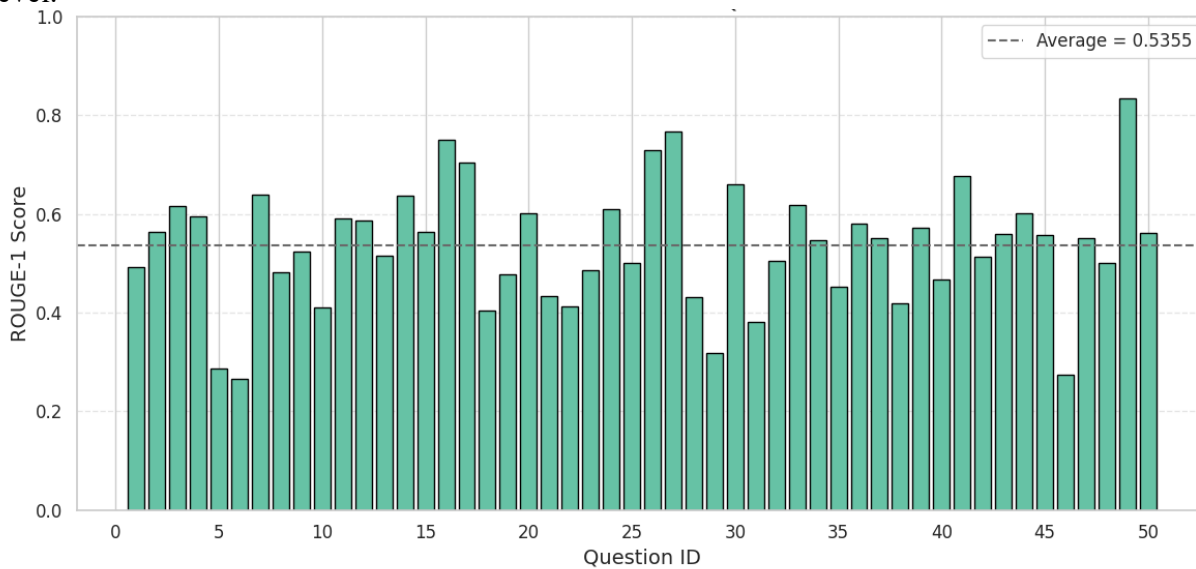

**Figure 7 ROUGE-1 scores for Q-A LLM**

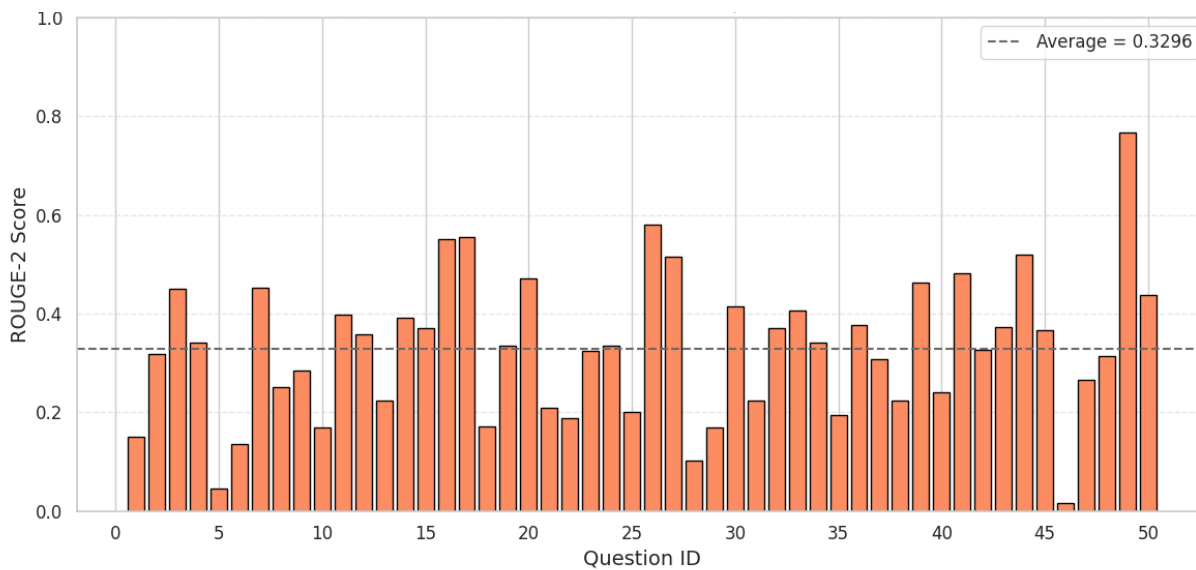

**Figure 8 ROUGE-2 scores for Q-A LLM**

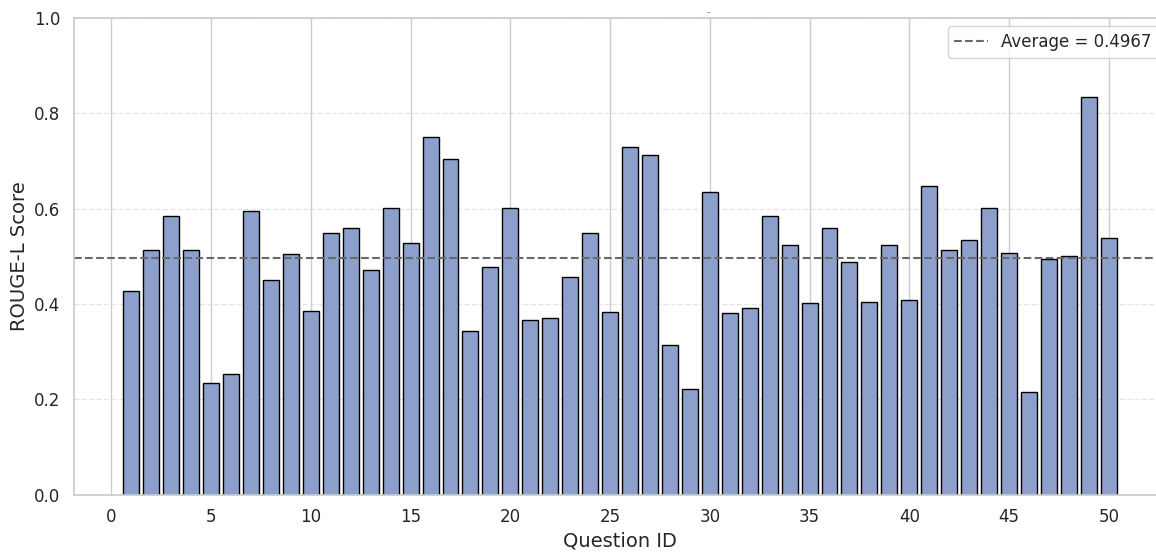

**Figure 9 ROUGE-L scores for Q-A LLM**

In addition to ROUGE, we evaluated the quality of the generated answers using token-level precision, recall, and accuracy metrics. **Figures 10, 11, and 12** show the precision, recall and accuracy for all responses. This approach assesses how many of the tokens in the system's response matched those in the reference answers. "Precision" measures the proportion of words in the generated response that correctly match those in the reference. "Recall" quantifies the extent to which words from the reference text are successfully captured in the generated response. "Accuracy" reflects the proportion of overlapping words relative to the total unique words present in either the reference or the generated response." These metrics help quantify not just overlap but also the correctness and completeness of the answers at a finer granularity than ROUGE alone. The model achieved 0.52 precision, 0.60 recall, and 0.37 accuracy. These results suggest that the system successfully retrieved the majority of relevant content (high recall), but also included additional tokens that were not present in the reference answers (moderate precision). The overall token-level accuracy of 37% reflects the challenge of generating exact matches with reference text. These results suggest that the model tends to recover key facts but may paraphrase or share information not included in the reference texts, leading to some divergence from the reference wording. **Table 3** presents a summary of the automatic evaluation metrics used to assess the LLM's performance.

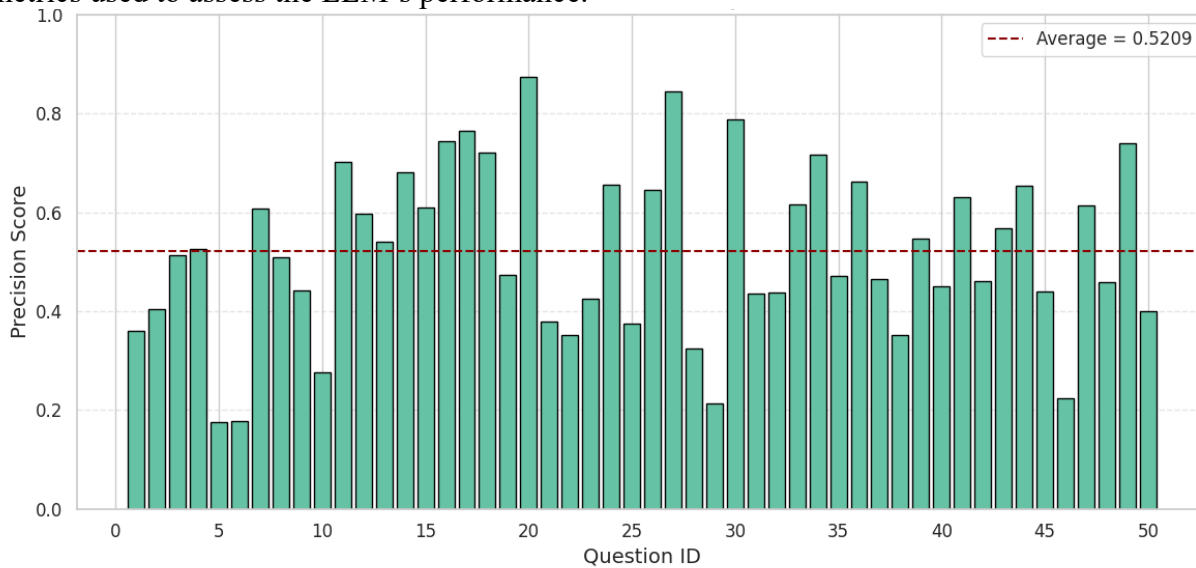

**Figure 10 Precision for Q-A LLM**

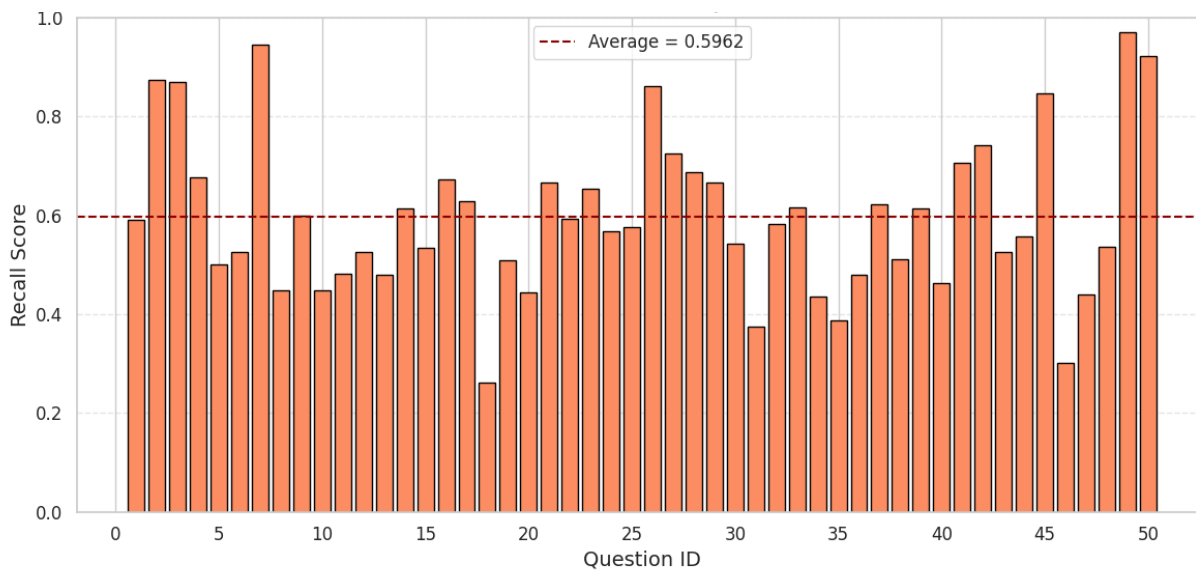

**Figure 11 Recall for Q-A LLM**

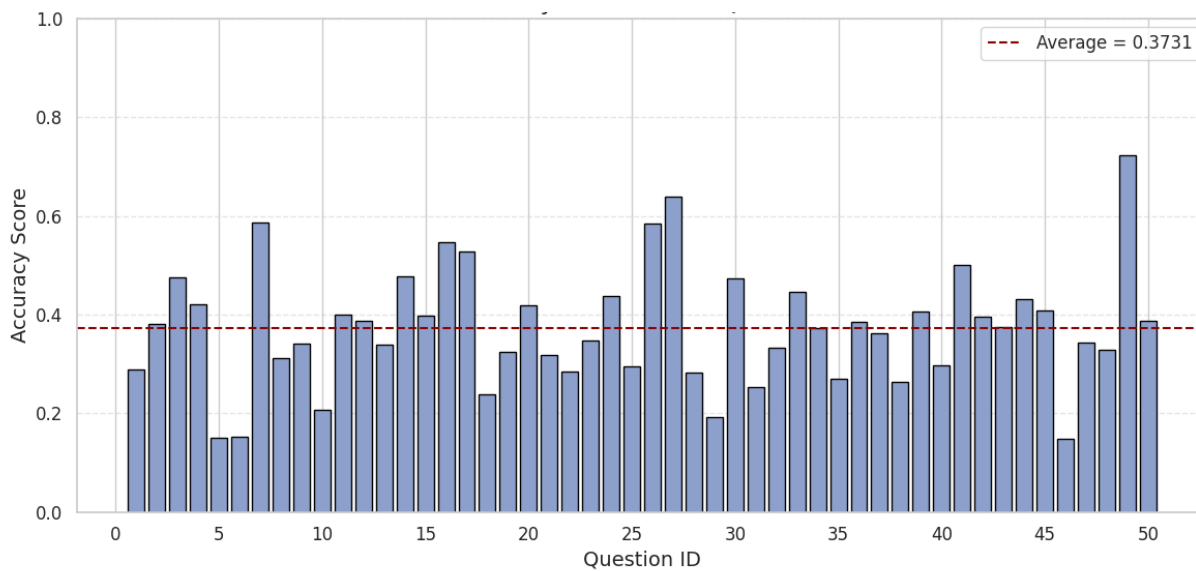

**Figure 12 Accuracy for Q-A LLM**

**TABLE 3 Summary of Automatic Q-A Evaluation Metrics**

| Metric | Performance | Comments |
|---|---|---|
| **Average ROUGE-1** | 0.54 | Good unigram overlap; relevant content captured |
| **Average ROUGE-2** | 0.33 | Lower phrase-level alignment; some structural drift |
| **Average ROUGE-L** | 0.50 | Strong sequential match; follows reference structure |
| **Average Precision** | 0.52 | Moderate relevance; some extraneous information |
| **Average Recall** | 0.60 | High content coverage; key facts often recovered |
| **Average Accuracy** | 0.37 | Partial token match; divergence due to paraphrasing |

**Manual Inspection of Q-A Responses**

The responses from the Q-A system were manually inspected and compared with the ground-truth answers to assess their factual accuracy and completeness. The findings of this manual inspection are summarized in **Table 4**. Overall, the LLM produced informative, coherent, and largely accurate responses, consistently going beyond simply answering the question to provide additional relevant details about the incidents. As noted earlier, the ground-truths in this study were derived primarily from the incident descriptions. The LLM correctly answered 49 out of 50 questions. The only incorrect response was to the question, "Which cities were affected by the railway cyberattack on Belarus in 2022?" While the LLM accurately described the broader incident, it stated, "The specific cities affected by this cyberattack were not detailed in the available information," despite such details being present in the dataset. This reflects a broader trend in which the model appeared to rely heavily on the "victim: country" field when identifying locations, occasionally overlooking more specific geographic information available in the "description" field. Despite this, the responses generally provided concise and accurate incident summaries while addressing the core of each question. However, in 16 cases, some relevant information was omitted (such as affected cities, technical execution details, or additional countries impacted) despite being present in the ground-truth. Conversely, in 10 cases, the LLM produced accurate and informative responses with information that was not present in the ground-truth answers, often by drawing on additional fields such as "attack type" or synthesizing relevant details from multiple related entries that may have been overlooked during ground-truth creation. These findings highlight the model's strengths in contextual understanding and synthesis, while also revealing areas for refinement in location specificity and detail coverage.

**TABLE 4 Manual Inspection of RAG Q-A Responses**

| Metric | Count | Comments |
|---|---|---|
| **Questions answered correctly** | 49 | Strong factual accuracy; most answers aligned with ground-truth |
| **Questions answered incorrectly** | 1 | Missed cities mentioned in description; rest of incident correctly captured |
| **Responses missing ground-truth details** | 16 | Omitted fine-grained information (e.g., cities, attack specifics, secondary countries, victim names) |
| **Responses containing additional details not in ground-truth** | 10 | Added relevant context from other entry components or multiple entries; more comprehensive than ground-truth |

**CONCLUSIONS**

This paper presents a novel, domain-specific approach for enhancing cyber-awareness in the transportation industry by utilizing a fine-tuned GPT-3.5 Turbo model, RAG, and GPT-4o mini. This system is used to build a structured database of transportation-related cyber incidents and answer questions about past transportation-related incidents. In our evaluation, the LLM-assisted classification process accurately identified the transportation mode in 82 out of 90 incidents. It successfully retained 74 out of 75 relevant incidents, demonstrating the model's effectiveness in structured data filtering. The observed accuracy and retention rates indicate that the filter effectively distinguishes between transportation-related and non-transportation-related incidents, supporting the construction of the transportation-specific database. The Q-A component achieved average scores of 0.54 (ROUGE-1), 0.33 (ROUGE-2), 0.50 (ROUGE-L), 0.52 precision, 0.60 recall, and 0.37 accuracy. These figures show that the system retrieves and generates responses that broadly reflect the core information of reference answers, as reflected in the relatively high recall. However, the moderate precision and low token-level accuracy indicate that answers may include additional context or deviate from expected phrasing. Manual inspection of the question-answering system confirmed its strong factual accuracy, with 49 out of 50 questions answered correctly. The LLM consistently exceeded the ground-truth responses by incorporating relevant details from other fields and multiple entries within the dataset, demonstrating its capacity to synthesize information across sources. However, in 16 responses, it omitted specific details such as impacted cities, methods of attack execution, or secondary countries affected, likely due to a reliance on higher-level fields like "victim: country" rather than more descriptive narrative fields. These findings highlight both the model's strength in contextual understanding and its limitations in consistently retrieving some fine-grained details.

While the system shows promise, several limitations must be acknowledged. For the classification pipeline, a clearer understanding of false negatives is needed. Because exhaustively validating the transportation relevance of all incidents in each dataset would require manual review of thousands of entries, we cannot conclusively determine the recall of the filter beyond the sampled test set. Additionally, the model struggled most with multimodal incidents, likely due to their underrepresentation in the training data, where most entries involved only a single transportation mode. A further limitation arises from the variability in incident descriptions across datasets. Some records contain only brief summaries, which constrains both the reference answers and the model's ability to generate concise responses. As a result, the LLM occasionally produces answers that are longer or more elaborate than the ground-truth. Despite these limitations, this work provides an important step toward the development of practical, LLM-powered tools for cybersecurity awareness in the transportation domain.

**FUTURE WORK**

This study holds significant potential for advancing cybersecurity awareness within the transportation sector. Some future directions to enhance and expand the existing framework to better achieve this objective have been identified, with an emphasis on increasing accessibility, incorporating automation, and improving usability. This work was motivated by initial challenges encountered while attempting to collect transportation-specific cyber incident data for the TraCR dataset. These improvements aim to simplify the process of accessing such information for researchers, practitioners, and policymakers alike.

Given that the overarching goal is to promote cybersecurity awareness in transportation, it would prove beneficial to make the system publicly available through a user-friendly online interface. This platform will provide stakeholders in both the public and private transportation sectors with better access to information regarding cyberattacks relevant to their domains. The website could include a searchable and filterable database of transportation-related cyberattacks, enabling users to explore incidents by mode, region, or time period. Additionally, an interactive dashboard could be incorporated to display key statistics and trends, offering a high-level overview of the current cybersecurity landscape in transportation. This would allow users to obtain rapid situational awareness without needing to submit specific queries.

Currently, the knowledge base used by the system is static and requires manual integration with the question answering approach based on retrieval-augmented generation. However, because new cyber incidents occur frequently, maintaining an up-to-date knowledge base is essential. A natural extension would involve automating this process by implementing regular web scraping from curated sources, including cyber incident databases and security news platforms. The collected data could then be passed through the existing filtering approach to retain only transportation-relevant incidents. By connecting the filtering system to the question answering module, the system could maintain an up-to-date knowledge base and deliver responses based on the most current information available, thereby enhancing its utility for real-time situational awareness.

**ACKNOWLEDGMENTS**

This work is based upon the work supported by the National Center for Transportation Cybersecurity and Resiliency (TraCR) (a U.S. Department of Transportation National University

## AUTHOR CONTRIBUTIONS

The authors confirm contribution to the paper as follows: study conception and design: O. Thomas, M. B. Munir; data collection: O. Thomas, M. B. Munir; analysis and interpretation of results: O. Thomas, M. B. Munir. J. Tine; draft manuscript preparation: O. Thomas, M. B. Munir, J. Tine, M. Rahman, Y. Cai, K. Akbar, M. N. Uddin, L. Khan, T. Hockstad, M. Chowdhury. All authors reviewed the results and approved the final version of the manuscript.

## DECLARATION OF CONFLICTING INTERESTS

The authors declared no potential conflicts of interest with respect to the research, authorship, and/or publication of this article.

## FUNDING

The authors disclosed receipt of the following financial support for the research, authorship, and/or publication of this article: This research was supported by the National Center for Transportation Cybersecurity and Resiliency (TraCR) (a U.S. Department of Transportation National University Transportation Center) headquartered at Clemson University, Clemson, South Carolina, USA, under Grants 69A3552344812 and 69A3552348317.

**REFERENCES**

1. Cost of a Data Breach 2024 | IBM. https://www.ibm.com/reports/data-breach. Accessed Jul. 18, 2025.

2. Maritime Cyber Attacks Increase By 900% In Three Years | Hellenic Shipping News Worldwide. https://www.hellenicshippingnews.com/maritime-cyber-attacks-increase-by-900-in-three-years/. Accessed Jul. 18, 2025.

3. Industry Spotlight: Port and Maritime. *Surefire Cyber*. https://www.surefirecyber.com/industry-spotlight-port-and-maritime/. Accessed Jul. 18, 2025.

4. Sivesind, C. Cyber Attacks on Railway Systems Increase by 220%. https://www.secureworld.io/industry-news/railway-cyber-attacks. Accessed Jul. 18, 2025.

5. Pangarkar, T. Automotive Cyber Security Statistics and Facts (2025). Market.us Scoop, Jan 14, 2025.

6. Aviation Sector Sees 600% Year-on-Year Increase in Cyberattacks | Thales Group. https://www.thalesgroup.com/en/worldwide/aerospace/press_release/aviation-sector-sees-600-year-year-increase-cyberattacks. Accessed Jul. 18, 2025.

7. Cybersecurity Dominates Concerns among the C-Suite, Small Businesses and the Nation | IBM. https://www.ibm.com/think/insights/cybersecurity-dominates-concerns-c-suite-small-businesses-nation. Accessed Jul. 18, 2025.

8. Coker, J. 95% of Data Breaches Tied to Human Error in 2024. *Infosecurity Magazine*. https://www.infosecurity-magazine.com/news/data-breaches-human-error/. Accessed Jul. 18, 2025.

9. Thomas, O., M. S. Salek, J.-M. Tine, M. Rahman, T. Hockstad, and M. Chowdhury. Cybersecurity in Transportation Systems: Policies and Technology Directions. http://arxiv.org/abs/2501.05356. Accessed Jul. 18, 2025.

10. Iain. How Can the Logistics and Transport Sector Combat Rising Threats? Transportation and Logistics International, Feb 26, 2025.

11. Cyber Events Database | Center for International and Security Studies at Maryland. https://cissm.umd.edu/cyber-events-database. Accessed Jun. 11, 2024.

12. The European Repository of Cyber Incidents. *EuRepoC: European Repository of Cyber Incidents*. https://eurepoc.eu/. Accessed Jun. 11, 2024.

13. NVD - Home. https://nvd.nist.gov/. Accessed Jun. 10, 2025.

14. ENISA Threat Landscape 2023 | ENISA. https://www.enisa.europa.eu/publications/enisa-threat-landscape-2023. Accessed Jun. 10, 2025.

15. Home Page | CISA. https://www.cisa.gov/. Accessed Aug. 1, 2025.

16. Kumar, A., R. Sharma, and P. Bedi. Towards Optimal NLP Solutions: Analyzing GPT and LLaMA-2 Models Across Model Scale, Dataset Size, and Task Diversity. *Engineering, Technology & Applied Science Research*, Vol. 14, No. 3, 2024, pp. 14219–14224. https://doi.org/10.48084/etasr.7200.

17. Berkane, T., M.-L. Charpignon, and M. Majumder. LLM-Based Web Data Collection for Research Dataset Creation. 2025.05.23.25328249. https://www.medrxiv.org/content/10.1101/2025.05.23.25328249v1. Accessed Jul. 28, 2025.

18. Xu, H., S. Wang, N. Li, K. Wang, Y. Zhao, K. Chen, T. Yu, Y. Liu, and H. Wang. Large Language Models for Cyber Security: A Systematic Literature Review. http://arxiv.org/abs/2405.04760. Accessed Jul. 28, 2025.

19. Salek, M. S., M. Chowdhury, M. B. Munir, Y. Cai, M. I. Hasan, J.-M. Tine, L. Khan, and M. Rahman. A Large Language Model-Supported Threat Modeling Framework for Transportation Cyber-Physical Systems. http://arxiv.org/abs/2506.00831. Accessed Jul. 28, 2025.

20. Zhao, C., G. Agrawal, T. Kumarage, Z. Tan, Y. Deng, Y.-C. Chen, and H. Liu. Ontology-Aware RAG for Improved Question-Answering in Cybersecurity Education. http://arxiv.org/abs/2412.14191. Accessed Jul. 28, 2025.

21. Arikkat, D. R., A. M., N. Binu, P. M., N. Biju, K. S. Arunima, V. P, R. Rehiman K. A., and M. Conti. IntellBot: Retrieval Augmented LLM Chatbot for Cyber Threat Knowledge Delivery. Presented at the 2024 IEEE 16th International Conference on Computational Intelligence and Communication Networks (CICN), 2024.

22. Significant Cyber Incidents | Strategic Technologies Program | Center for Strategic and International Studies (CSIS). https://www.csis.org/programs/strategic-technologies-program/significant-cyber-incidents. Accessed May 29, 2024.

23. Maritime Cyber Attack Database (MCAD) | NHL Stenden University of Applied Sciences. https://www.nhlstenden.com/en/maritime-cyber attack-database. Accessed Jun. 11, 2024.

24. Hsu, H.-L., and J. Tzeng. DAT: Dynamic Alpha Tuning for Hybrid Retrieval in Retrieval-Augmented Generation. http://arxiv.org/abs/2503.23013. Accessed Jul. 29, 2025.

25. Sawarkar, K., A. Mangal, and S. R. Solanki. Blended RAG: Improving RAG (Retriever-Augmented Generation) Accuracy with Semantic Search and Hybrid Query-Based Retrievers. Presented at the 2024 IEEE 7th International Conference on Multimedia Information Processing and Retrieval (MIPR), 2024.

26. BM25 Retriever - LlamaIndex. https://docs.llamaindex.ai/en/stable/examples/retrievers/bm25_retriever/. Accessed Jun. 5, 2025.

27. Mastering ROUGE Matrix: Your Guide to Large Language Model Evaluation for Summarization with Examples. *DEV Community*. https://dev.to/aws-builders/mastering-rouge-matrix-your-guide-to-large-language-model-evaluation-for-summarization-with-examples-jjg. Accessed Jul. 18, 2025.